\pdfoutput=1
\RequirePackage{ifpdf}
\ifpdf 
\documentclass[pdftex]{sigma}
\else
\documentclass{sigma}
\fi

\begin{document}

\allowdisplaybreaks

\renewcommand{\thefootnote}{$\star$}

\renewcommand{\PaperNumber}{070}

\FirstPageHeading

\ShortArticleName{Superintegrable Extensions of Superintegrable Systems}

\ArticleName{Superintegrable Extensions\\ of Superintegrable Systems\footnote{This
paper is a contribution to the Special Issue ``Superintegrability, Exact Solvability, and Special Functions''. The full collection is available at \href{http://www.emis.de/journals/SIGMA/SESSF2012.html}{http://www.emis.de/journals/SIGMA/SESSF2012.html}}}

\Author{Claudia M.~CHANU~$^\dag$, Luca DEGIOVANNI~$^\ddag$ and Giovanni RASTELLI~$^\S$}

\AuthorNameForHeading{C.M.~Chanu, L.~Degiovanni and G.~Rastelli}

\Address{$^\dag$~Dipartimento di Matematica, Universit\`a di Torino,\\
\hphantom{$^\dag$}~Torino, via Carlo Alberto 10, Italy}
\EmailD{\href{mailto:claudiamaria.chanu@unito.it}{claudiamaria.chanu@unito.it}}

\Address{$^\ddag$~Formerly at Dipartimento di Matematica, Universit\`a di Torino,\\
\hphantom{$^\ddag$}~Torino, via Carlo Alberto 10, Italy}
\EmailD{\href{mailto:luca.degiovanni@gmail.com}{luca.degiovanni@gmail.com}}

\Address{$^{\S}$~Independent researcher, cna Ortolano 7, Ronsecco, Italy}
\EmailD{\href{mailto:francesco.calogero@roma1.infn.it}{giorast.giorast@alice.it}}

\ArticleDates{Received July 30, 2012, in f\/inal form September 27, 2012; Published online October 11, 2012}

\Abstract{A procedure to extend a superintegrable system into a new superintegrable one is systematically tested for the known systems on $\mathbb E^2$ and $\mathbb S^2$ and for a family of systems def\/ined on constant curvature manifolds. The procedure results ef\/fective in many cases including Tremblay--Turbiner--Winternitz and three-particle Calogero systems.}

\Keywords{superintegrable Hamiltonian systems; polynomial f\/irst integrals}

\Classification{70H06; 70H33; 53C21}

\renewcommand{\thefootnote}{\arabic{footnote}}
\setcounter{footnote}{0}

\section{Introduction}

Given a natural Hamiltonian $L$ with $n$ degrees of freedom, satisfying some additional geometric conditions, it is shown in~\cite{sigma11} how to generate a $n+1$ degrees of freedom Hamiltonian~$H$, called the extension of~$L$ and depending on an integer parameter $m\in \mathbb N\setminus\{0\}$, such that $H$
admits two new  independent f\/irst integrals: $H$~itself and a polynomial in the momenta of degree~$m$. This implies that, if  $L$ is superintegrable with $2n-1$ independent f\/irst integrals, then all the extended Hamiltonians~$H$ also are superintegrable for any~$m$ with the maximal number of $2n+1=2(n+1)-1$  f\/irst integrals, one of them of arbitrary degree~$m$ whose expression is explicitly computed by means of a simple iterative process. The extension procedure, summarized in Section~\ref{section2}, is applied to the superintegrable systems on~$\mathbb E^2$ (Section~\ref{section3}) and $\mathbb S^2$ (Section~\ref{section5}) as listed in~\cite{KM?}. These are all the superintegrable systems on~$\mathbb S^2$ and~$\mathbb E^2$ admitting two independent f\/irst integrals of degree two in the momenta other than the Hamiltonian. It is found that a~great part of them  admits superintegrable extensions and for some of them the extensions are explicitly determined.
In Section~\ref{section4} the possible natural Hamiltonians admitting an extension are determined on~$\mathbb E^n$ and the superintegrable systems of Calogero and Wolfes are considered as examples in~$\mathbb E^3$.
In Section~\ref{section6} the extension procedure is applied to a class of Hamiltonians on constant curvature manifolds to which some generalizations of the Tremblay--Turbiner--Winternitz (TTW) system belong ~\cite{TTW}. These generalizations of the TTW systems are superintegrable for rational values of certain parameters, for which they admit polynomial f\/irst integrals of degree greater than two~\cite{KKM, MPY}. The cases allowing the extensions are determined and the TTW systems are among them, obtaining in this way their superintegrable extensions. Other superintegrable generalizations in higher dimensions of the TTW system are obtained in~\cite{KKMCu}.

\section{Extensions of superintegrable systems}\label{section2}

We resume in the following statement the main results proved in \cite{sigma11}.

\begin{theorem} \label{theorem1}
Let  $Q$ be a $n$-dimensional $($pseudo-$)$Riemannian manifold with metric tensor $\mathbf g$.
The natural Hamiltonian $L=\frac{1}{2} g^{ij}p_ip_j+V(q^i)$ on $M=T^*Q$ with canonical coordinates $(p_i,q^i)$ admits an extension
\begin{gather}\label{HamExt}
H=\frac{1}{2} p_u^2+\alpha(u)L + f(u)
\end{gather}
with a first integral $F=U^m(G)$ where
\begin{gather*}
U=p_u+ \gamma(u) X_L,
\end{gather*}
$X_L$ is the Hamiltonian vector field of $L$ and $G(q^i)$,  if and only if the following conditions hold:
\begin{enumerate}\itemsep=0pt
\item[$i)$]
the functions $G$ and $V$ satisfy
\begin{gather}\label{HessTeo}
 \mathbf{H}(G) + mc  \mathbf{g}G=\mathbf 0, \qquad m\in {\mathbb N} \setminus \{0\}, \qquad c\in \mathbb R,
\\
\label{VTeo} \nabla V \cdot \nabla G-2m(cV+L_0)G=0,\qquad L_0\in \mathbb R,
\end{gather}
where $\mathbf H(G)_{ij}=\nabla_i\nabla_jG$ is the Hessian tensor of~$G$.
\item[$ii)$] for $c=0$ the extended Hamiltonian $H$ is
\begin{gather}\label{Ext0}
H=\frac{1}{2}p_u^2+mA(L+V_0)+mL_0A^2(u+u_0)^2,
\end{gather}
for $c\neq 0$ the extended Hamiltonian $H$ is
\begin{gather}\label{Extk}
H=\frac{1}{2}p_u^2+\frac{m(cL+L_0)}{S^2_\kappa(cu+u_0)}+W_0,
\end{gather}
with $\kappa$, $u_0$, $V_0$, $W_0$, $A\in \mathbb R$, $A\neq0$ and
\begin{gather*}
S_\kappa(x)= \begin{cases}
\dfrac{\sin\sqrt{\kappa}x}{\sqrt{\kappa}}, & \kappa>0, \\
x, & \kappa=0, \\
\dfrac{\sinh\sqrt{|\kappa|}x}{\sqrt{|\kappa|}}, & \kappa<0.
\end{cases}
\end{gather*}
\end{enumerate}
\end{theorem}

Dynamically, extended Hamiltonians (\ref{HamExt}) can be written  as
\begin{gather*}
\frac 12 p_u^2-\dfrac m{S_\kappa^2(cu+u_0)}\eta-h =0,\qquad
cL+L_0+\eta=0,
\qquad \mbox{if} \quad c\neq 0,
\end{gather*}
and
\begin{gather*}
\frac 12 p_u^2+mL_0A^2u^2- \eta-h =0,\qquad
mA(L+V_0)+\eta=0,
\qquad \mbox{if}  \quad c = 0,
\end{gather*}
with $H=h$, and where constant $\eta$ can be understood either as a separation constant, between the elements of $H$ depending on $(u,p_u)$ and those depending on $(q^i,p_i)$, or as a coupling constant merging together the Hamiltonian $L$ on $T^*Q$ and the Hamiltonian
\begin{gather*}
\frac 12 p_u^2-\dfrac m{S_\kappa^2(cu+u_0)},\qquad c\neq 0,
\qquad \mbox{or}
\qquad
\frac 12 p_u^2+mL_0A^2u^2, \qquad c=0,
\end{gather*}
depending on $(u,p_u)$, to build the extended Hamiltonian $H$. Several examples  are given in~\cite{sigma11}.

Under the hypotheses of Theorem~\ref{theorem1}, recalling that $C_\kappa(x)=\frac d{dx}S_\kappa(x)$, the polynomials in the momenta
\begin{gather*}
U^mG= \begin{cases}
\left(p_u+\dfrac {C_\kappa (cu+u_0)}{S_\kappa (cu+u_0)}X_L\right)^mG, & c\neq 0, \\
\left(p_u-A(u+u_0)X_L\right)^mG, & c=0,
\end{cases}
\end{gather*}
are f\/irst integrals of $H$ of degree $m$. For example,
\begin{gather*}
UG=Gp_u+\gamma(u)X_L(G)=Gp_u+\gamma(u)\{L,G\},
\end{gather*}
where $\{\,,\, \}$ are the Poisson brackets.
Another way to calculate $U^mG$ is to apply the formula~\cite{CDRG}
\begin{gather*}
U^mG=P_mG+D_mX_LG,
\end{gather*}
with
\begin{gather*}
P_m=\sum_{k=0}^{[m/2]}{\left( \begin{matrix} m \cr 2k \end{matrix} \right) \gamma^{2k}p_u^{m-2k}\left(-2m(cL+L_0)\right)^k},
\\
D_m=\sum_{k=0}^{[m/2]-1}{\left( \begin{matrix} m \cr 2k+1 \end{matrix} \right) \gamma^{2k+1}p_u^{m-2k-1}\left(-2m(cL+L_0)\right)^k}, \qquad m>1,
\end{gather*}
where $[\cdot]$ denotes the integer part, $D_1=\gamma$ and
\begin{gather}\label{eqg}
\gamma(u)= \begin{cases}
\dfrac{C_\kappa(cu+u_0)}{S_\kappa(cu+u_0)}, & c\neq0, \\
-A(u+u_0), & c=0.
\end{cases}
\end{gather}

\begin{remark}\label{remark1}
For $c=0$, if $L_0=0$ then $p_u$ is a f\/irst integral of the extended Hamiltonian and the extended potential is merely~$mAV$, with~$m$ and~$A$ constants. Therefore the extension is trivial. Otherwise, a harmonic oscillator term in the variable $u$, attractive or repulsive, is added to the potential~$mAV$. In the case $c\neq 0$, we remark that, in the extended Hamiltonian~(\ref{Extk}), the potential $V$ is multiplied by a non-constant factor depending on~$u$.
\end{remark}

\begin{remark} \label{remark2}
The expression  (\ref{eqg}) of $\gamma(u)$ is determined  in \cite{sigma11} as the general solution  of the dif\/ferential equation
\begin{gather}\label{deq}
\gamma '+c(\gamma^2 +\kappa)=0,
\end{gather}
for real values of $\gamma$, $u$, $c$ and $\kappa$. However, the general solution of equation (\ref{deq}) in the complex case can be considered too. Its expression is the same as (\ref{eqg}) if we extend functions $S_\kappa$ and $C_\kappa$ to complex values of~$x$ and~$\kappa$ by using the standard exponential expressions of trigonometric functions.
After this generalization, the extension procedure characterized by Theorem~\ref{theorem1} can be applied also to the complex case.
\end{remark}

In \cite{sigma11} it is proved that, if $n>1$, then~(\ref{HessTeo}) admits a complete solution $G$ depending on a maximal number of parameters~$(a_i)$, with $i=0, \ldots,n$, if\/f the sectional curvature of~$Q$ is constant and equal to~$mc$. Once such a solution~$G$ is known, an extension of~$L$ is possible if\/f the compatibility condition~(\ref{VTeo}) on~$V$ is satisf\/ied. A sharper result on constant curvature manifolds is the following

\begin{proposition}\label{prop}
On a  manifold  with constant curvature~$K$ the only eigenvalues~$mc$ for the Hessian equation \eqref{HessTeo} are either zero or the curvature~$K$. Moreover, if $K\neq 0$ and $mc=0$ then the only solutions of~\eqref{HessTeo} are $G=\operatorname{const}$.
\end{proposition}

\begin{proof}
The equation (\ref{HessTeo}), written in components, becomes
\begin{gather*}
\nabla_i\nabla_j G +mcg_{ij}G= \partial_{ij}G-\Gamma^k_{ij}\partial_k G+mcg_{ij}G= 0.
\end{gather*}
and the integrability conditions that each solution $G(q^i)$ must satisfy  are given by
(see \cite{sigma11} for details)
\begin{gather*}
R^k_{hij}z_k=mc(g_{jh}z_i-g_{ih}z_j), \qquad \forall\,  h,\quad \forall\,  i\neq j,
\end{gather*}
where $z_k={\partial_k G}/{G}$ and
$R^k_{hij}=\partial_i \Gamma_{jl}^k -\partial_j \Gamma_{il}^k +\Gamma_{jl}^h\Gamma_{ih}^k
- \Gamma_{il}^h\Gamma_{jh}^k$
is the Riemann tensor of the metric.
For a constant curvature manifold we have $R_{hlij}=K(g_{jl}g_{hi}-g_{il}g_{hj})$ and the above conditions become
\begin{gather*}
(K-mc)(g_{jh}z_i-g_{ih}z_j)=0, \qquad \forall\, h,\quad \forall\, i\neq j.
\end{gather*}
By choosing orthogonal coordinates, we see that, since $i\neq j$, the  equations are  identically satisf\/ied for $h\neq i,j$. Otherwise, they reduce to $(K-mc)g_{jj}z_i=0$. Hence, for  $mc\neq K$, the
only possibility is $z_i=0$  for all $i$ (that is, $G$ is a constant). For $mc\neq K$ and $mc\neq 0$, by~(\ref{HessTeo})  we get $G=0$, thus $mc$ is not an eigenvalue.
\end{proof}

In \cite{sigma11} it is shown that the f\/irst integral  $U^m(G)$ is functionally independent from $H$, $L$ and all its possible f\/irst integrals $L_i$ in $T^*Q$. It is straightforward to see that $L$ and $L_i$ are f\/irst integrals of $H$, therefore, if $L$ is a superintegrable Hamiltonian with $2n-1$ f\/irst integrals, inclu\-ding~$L$,  then $H$ is superintegrable too with $2n-1+2=2(n+1)-1$ f\/irst integrals including $H$ itself. It follows that the extension procedure applied to a superintegrable Hamiltonian~$L$, under the hypothesis of Theorem~\ref{theorem1}, always produces a new superintegrable Hamiltonian~$H$.
Given a~superintegrable system of Hamiltonian $L=\frac 12 g^{ij}p_ip_j+V(q^i)$ with a  conf\/iguration manifold~$Q$ of constant curvature $K$, its extension to another superintegrable system of Hamiltonian~$H$, when possible, can be obtained  by applying the following algorithm (see also~\cite{sigma11}):
\begin{enumerate}\itemsep=0pt
\item [1.] Solve equation (\ref{HessTeo}) on the manifold $Q$, with $c=K/m$, to compute the general form of the  function $G(q^i,a_0,\ldots,a_n)$.
\item [2.] Solve equation (\ref{VTeo}) with the given $V$ for some of the parameters $(a_i)$ in $G$. This is a crucial step, because if no solution  is found, except for the trivial one  $G=0$, then the extension is not possible.
\item [3.] Determine the extension via Theorem~\ref{theorem1}.
\item [4.] Compute $U^m(G)$  to obtain the additional f\/irst integral.
\end{enumerate}
In the following sections we analyze several examples of extensions of superintegrable systems.

\section{Superintegrable extensions of $\mathbb E^2$ systems}\label{section3}
Since the curvature of $\mathbb E^2$ is zero, equation~(\ref{HessTeo}) admits, by Proposition \ref{prop}, solutions $G\neq 0$ only for  $mc=0$. Thus, we are in the case $c=0$ of Theorem~\ref{theorem1} and the extended Hamiltonian $H$ is in the form~(\ref{Ext0}). In~\cite{sigma11} the complete solution~$G$ of~(\ref{HessTeo})  is computed, in standard Cartesian coordinates of~$\mathbb E^2$, as
\begin{gather}\label{ge1}
G=a_0+ a_1x+a_2y.
\end{gather}
Equation (\ref{VTeo}) becomes here
\begin{gather*}
\partial_x V \partial_x G+\partial_y V \partial_y G=2mL_0G,
\end{gather*}
whose general solution is
\begin{gather}\label{VE2}
V=mL_0\big[\big(x+x^0\big)^2+\big(y+y^0\big)^2\big]+F(a_2x-a_1y),
\end{gather}
with the constraint $a_0= a_{1}x^0+a_{2}y^0$,  where $x^0,y^0\in \mathbb R$ or $\mathbb C$ and $F$ is any regular function of the argument. The extension, after the non restrictive assumptions $V_0=u_0=0$ and $A=m^{-1}$, becomes
\begin{gather*}
H=\frac12 p_u^2+L+\frac{L_0}{m}u^2=\frac12\big(p_u^2+p_x^2+p_y^2\big)+V+\frac{L_0}{m}u^2.
\end{gather*}
In~\cite{KM?} the list of all superintegrable potentials in $\mathbb E^2$ with three independent quadratic in the momenta f\/irst integrals is given, up to isometries and ref\/lections. In that article $\mathbb E^2$ is assumed to be a two-dimensional complex manifold. According to Remark~\ref{remark2} we apply in this case the same procedure developed for the real case and allow all functions, variables and parameters (except for $m\in \mathbb{N}$) to take indif\/ferently real or complex values, in this one and all the following sections.
By setting $z=x+iy$, $\bar z=x-iy$ (remark that, despite the notation, if $x$ and $y$ are complex coordinates, then $z$ and $\bar z$ are not complex conjugate one of the other), the list is
\begin{alignat*}{3}
&E1 \ \ \ && V=\frac {\alpha_1}{x^2}+ \frac {\alpha_2}{y^2} +\alpha_3 \big(x^2+y^2\big), &\\
&E2\ \ \ && V=\alpha_1 x+ \frac {\alpha_2}{y^2}+ \alpha_3 \big(4x^2+y^2\big), &\\
&E3\ \ \ && V=\alpha_3 \big(x^2+y^2\big), &\\
&E4\ \ \ && V=\alpha_1(x+iy), &\\
&E5\ \ \ && V=\alpha_1 x, &\\
&E6\ \ \ && V=\frac {\alpha_1} {x^2}, &\\
&E7\ \ \ && V=\frac {\alpha_1 \bar z}{\sqrt{\bar z^2-k^2}}+\frac{\alpha_2 z}{\sqrt{\bar z^2-k^2}(\bar z+\sqrt{\bar z^2-k^2})^2}+\alpha_3 z\bar z, &\\
&E8\ \ \ && V=\frac {\alpha_1 z}{\bar z^3}+\frac {\alpha_2}{\bar z^2}+\alpha_3 z\bar z, &\\
&E9\ \ \ && V=\frac {\alpha_1}{\sqrt{\bar z}}+\alpha_2 x+\alpha_3\frac {x+\bar z}{\sqrt{\bar z}}, &\\
&E10\ \ \ && V=\alpha_1 \bar z+\alpha_2 \left(z-\frac 32 \bar z^2\right)+ \alpha_3 \left(z \bar z -\frac 12 \bar z^3\right), &\\
&E11\ \ \ && V= \alpha_1 z+\frac {\alpha_2 z}{\sqrt{\bar z}}+\frac {\alpha_3}{\sqrt {\bar z}}, &\\
&E12\ \ \ && V=\frac {\alpha_1 \bar z}{\sqrt{\bar z^2+k^2}}, &\\
&E13\ \ \ && V=\frac  {\alpha_1}{\sqrt{ \bar z}}, &\\
&E14\ \ \ && V=\frac {\alpha_1} {\bar z^2}, &\\
&E15\ \ \ && V=h(\bar z),\quad \text{for any function} \ h, &\\
&E16\ \ \ && V=\frac 1{\sqrt{x^2+y^2}}\left(\alpha_1+\frac {\alpha_2} {x+\sqrt{x^2+y^2}}+\frac{\alpha_3} {x-\sqrt{x^2+y^2}}\right), &\\
&E17\ \ \ && V=\frac {\alpha_1} {\sqrt{z \bar z}}+\frac {\alpha_2}{z^2}+\frac {\alpha_3}{z\sqrt{z \bar z}}, &\\
&E18\ \ \ && V=\frac {\alpha_1}{\sqrt{x^2+y^2}}, &\\
&E19\ \ \ && V=\frac {\alpha_1 \bar z}{\sqrt{\bar z^2-4}}+\frac {\alpha_2} {\sqrt{z(\bar z+2)}}+\frac {\alpha_3} {\sqrt{z(\bar z-2)}}, &\\
&E20\ \ \ && V=\frac 1{\sqrt{x^2+y^2}}\left(\alpha_1+\alpha_2 \sqrt{x+\sqrt{x^2+y^2}}+ \alpha_3 \sqrt{x-\sqrt{x^2+y^2}}\right), &
\end{alignat*}
with $(\alpha_i)$, $k\in \mathbb C$. The equation of compatibility~(\ref{VTeo}) is  equivalent to a linear homogeneous expression in~$(a_i)$ with coef\/f\/icients linear but not homogeneous in $\alpha_i$. This expression vanishes only for some suitable choices of the parameters~$a_i$ and~$\alpha_i$,  depending on the extension parame\-ters~$m$,~$L_0$. The solution~(\ref{VE2}) of~(\ref{VTeo}) shows that  a non null function $G$ of the form~(\ref{ge1}) satisfying~(\ref{VTeo}) exists only for the following potentials

\begin{center}
\begin{tabular}{|c|l|c|c|}
\hline
& \hspace{15mm} $V$ & $G$ & particular cases of \\
\hline
$i$ &$mL_0\big(x^2+y^2\big)$\tsep{3pt}\bsep{4pt} &  $a_1x+a_2y$ & $E1$, $E3$, $E7$, $E8$ \\
$ii$ &$\dfrac{\alpha_1}{x^2}+mL_0\big(x^2+y^2\big)$\bsep{8pt} & $a_2y$ & $E1$ \\
$iii$ & $\dfrac{\alpha_2}{y^2}+mL_0\big(x^2+y^2\big)$\bsep{8pt} & $a_1x$ & $E1$ \\
$iv$ & $\alpha_1x+mL_0\big(4x^2+y^2\big)$\bsep{8pt} &  $a_2y$  & $E2$ \\
$v$ & $\alpha_1x+\dfrac {\alpha_2}{y^2}+\dfrac{mL_0}4\big(4x^2+y^2\big)$\bsep{8pt} & $a_1\left(\dfrac{\alpha_1}{2mL_0}+x\right)$ & $E2$\\
$vi$ & $\dfrac{\alpha_1\bar z}{\sqrt{\bar z^2-k^2}}+mL_0\big(x^2+y^2\big)$\bsep{8pt} & $a_1\bar z$ & $E7$\\
$vii$ & $\dfrac{\alpha_2}{\bar z^2}+mL_0\big(x^2+y^2\big)$\bsep{8pt} & $a_1\bar z$ & $E8$\\
$viii$ & $\alpha_1\bar z+\alpha_2\left(z-\dfrac 32\bar z^2\right)+ mL_0\left(z\bar z-\dfrac{\bar z^3}2\right)$
\bsep{8pt} & $a_1\left(\dfrac{\alpha_2}{mL_0}+\bar z\right)$ & $E10$\\
\hline
\end{tabular}
\end{center}

The potentials admitting $G(a_i)$ depending on several~$a_i$  allow the existence of dif\/ferent f\/irst integrals of the form $U^mG(a_i)$. However, since the system is already superintegrable by including a single~$U^mG(a_i)$,
all the other f\/irst integrals obtained in this way functionally depend on the known ones.

Because $L_0\neq 0$, a necessary condition for the extensibility is the presence of a harmonic term in the potential~$V$.

\subsection*{Extensions of the harmonic oscillators}

As an example, we analyze into details the extensions of the isotropic harmonic oscillator (a~particular case of $E1$, $E3$, $E7$, $E8$)
\begin{gather*}
V_i=\alpha_3\big(x^2+y^2\big),
\end{gather*}
and of the anisotropic one (a particular case of $E2$)
\begin{gather}\label{Va}
V_a=\alpha_3\big(4x^2+y^2\big).
\end{gather}
In both cases we assume $\alpha_3\neq0$ to avoid a trivial potential.

The only possible extension for $V_i$ is
\begin{gather}
H=\frac 12 \left(p_u^2+p_x^2+p_y^2\right)+\alpha_3\big(x^2+y^2\big)+\frac{L_0}{m}u^2\nonumber\\
\hphantom{H}{}
=\frac 12 \left(p_u^2+p_x^2+p_y^2\right)+\alpha_3\left(x^2+y^2+\frac{u^2}{m^2}\right),\label{extVi}
\end{gather}
with the constraint $\alpha_3-mL_0=0$ due to (\ref{VTeo}) that sets the potential in the tabulated form.
The Hamiltonian~(\ref{extVi}) represents an anisotropic oscillator in $\mathbb E^3$ and shows explicitly its superintegrability being $m$ an integer.
Recalling that $U=p_u-m^{-1}uX_L$, and by setting $X_LG=a_1p_x+a_2p_y=P$, we obtain as an example  the expression of~$U^4G$
\[
U^4G = Gp_u^4-up_u^3P-\dfrac 34 a_3Gu^2p_u^2+\dfrac{a_3}8u^3p_uP+\dfrac{a_3^2}{64}Gu^4.
\]

The anisotropic oscillator~(\ref{Va}) admits, instead, two extensions, corresponding to the two dif\/ferent functions $G$ and two dif\/ferent relations between $\alpha_3$ and $L_0$ (items $(iv)$ and $(v)$ of the table with $\alpha_1=\alpha_2=0$). In order to analyze this case in full generality, an iterated procedure of extension can be applied. By starting from the one-dimensional oscillator
\begin{gather*}
H_1=\frac12 p_1^2+\omega x_1^2,
\end{gather*}
we build a f\/irst extension
\begin{gather*}
H_2=\frac 12 p_2^2+H_1+\frac \omega{m_1^2}x_2^2=\frac 12\left(p_1^2+p_2^2\right)+\omega x_1^2+\frac {\omega}{m_1^2}x_2^2,
\end{gather*}
where we use $x_2=u$ and, because of (\ref{VTeo}), we have $G_1=a_1x_1$ and $L_0=\omega/m_1$. The resulting potential is an anisotropic oscillator and our procedure produces the third f\/irst integral of degree~$m_1$ in the momenta $U^{m_1}G_1=(p_2-m_1^{-1}X_{H_1})^{m_1}G_1$.  The potential of~$H_2$ coincides with~$V_a$ for $m_1=2$ and $\omega=4\alpha_3$ (this is the unique case  with a third quadratic f\/irst integral).

A further step is the research of an extension of~$H_2$
\begin{gather*}
H_3=\frac 12\left( p_1^2+p_2^2+p_3^2\right)+\omega x_1^2+\frac {\omega}{m_1^2}x_2^2+\frac{L_0}{m_2}x_3^2,
\end{gather*}
where the value of $L_0$ has to be determined.
In this case, since the general~$G_2$ is~(\ref{ge1}), condition~(\ref{VTeo}) becomes
\begin{gather*}
a_1\omega x_1+a_2\frac{\omega}{m_1^2}x_2=m_2L_0(a_0+a_1x_1+a_2x_2).
\end{gather*}
Hence, we get $a_0=0$, and
\begin{gather*}
a_1\left(\omega-m_2L_0\right)=0,\qquad
a_2\left(\dfrac{\omega}{m_1^2}-m_2L_0\right)=0.
\end{gather*}
If $a_1=0$, we need $a_2\neq 0$ and
\begin{gather*}
L_0=\frac{\omega}{m_1^2m_2}, \qquad G_2=a_2x_2.
\end{gather*}
If $a_1\neq 0$ and $a_2=0$, then we have
\begin{gather*}
L_0=\frac{\omega}{m_2}, \qquad G_2=a_1x_1.
\end{gather*}
Finally, if both $a_1\neq 0$ and $a_2\neq0$ the two conditions are satisf\/ied if\/f~$m_1^2=1$ (i.e., $H_2$ is the Hamiltonian of an isotropic oscillator) and
\begin{gather*}
L_0=\frac\omega{m_2}, \qquad G_2=a_1x_1+a_2x_2.
\end{gather*}
In the f\/irst two cases, $H_3$ represents an anisotropic harmonic oscillator, which is superintegrable because $m_1$, $m_2$ are integers, but they have dif\/ferent functions $G_2$ and consequently  dif\/ferent f\/irst integrals $U^{m_2}G_2=(p_3-m_2^{-1}X_{H_2})^{m_2}G_2$. If we set $m_1=2$, $m_2=m$ and $\omega=4\alpha_3$ in order to restrict ourselves to the potential~(\ref{Va}), we obtain two possible extensions: if $a_1=0$ we have the relation $\alpha_3-mL_0=0$ that gives the f\/irst form in the table. If, otherwise, $a_2=0$ we have the relation $4\alpha_3-mL_0=0$ that gives the second one.

The extension procedure can be iterated indef\/initely obtaining at the $n$-th step an $n$-dimen\-sio\-nal anisotropic oscillator  with a complete set of f\/irst integrals $(U^{m_1}G_1,U^{m_2}G_2,\ldots,U^{m_{n-1}}G_n)$ of degree $(m_1,m_2,\ldots, m_{n-1})$, that, together with the $n$ Hamiltonians $(H_1,H_2, \ldots, H_n)$  make the system superintegrable. We remark that the systems obtained in this way are characterized by the fact that the frequencies are all integer multiples of one of them. See~\cite{JH,EW} for additional details on superintegrability of anisotropic oscillators.

\section[Superintegrable extensions of $\mathbb E^n$]{Superintegrable extensions of $\boldsymbol{\mathbb E^n}$}\label{section4}

It is straightforward to generalize to $\mathbb E^n$ the procedure previously applied to~$\mathbb E^2$. Let us consider in~$\mathbb E^n$  with Cartesian coordinates the Hamiltonian
\begin{gather*}
L=\frac 12\big(p_1^2+p_2^2+\cdots +p_n^2\big)+V(x_1,x_2,\ldots ,x_n).
\end{gather*}
The general solution of (\ref{HessTeo}) and (\ref{VTeo}) are
\begin{gather}
 G=a_0+a_1x_1+a_2x_2+\cdots +a_{n}x_n \nonumber \\
 V=mL_0\big[\big(x_1+x_1^0\big)^2+\cdots +\big(x_n+x_n^0\big)^2\big]+F(a_1x_2-a_2x_1,\ldots,a_1x_n-a_{n}x_1), \label{VEn}
\end{gather}
with the constraint $a_0=\sum\limits_{i=1}^n a_{i}x_i^0$,  where $x_i^0\in \mathbb R$ or $\mathbb C$, $F$ is any regular function of the arguments and $L_0\neq 0$. The corresponding extension is
\begin{gather*}
H=\frac{1}{2}p_u^2+mA(L+V_0)+mL_0A^2(u+u_0)^2.
\end{gather*}
We remark that, as well as in dimension 2, the presence of a harmonic term in $V$ is a necessary condition for the extensibility.

\subsection*{Extensions of the three-body Calogero and Wolfes systems}
We consider the particular case of $n=3$. If $a_2=a_3=a_1$ and $F(X_1,X_2)$  in~(\ref{VEn}) is
\begin{gather*}
F=k\left(X_1^{-2}+X_2^{-2}+(X_1-X_2)^{-2}\right),
\end{gather*}
with $k\in \mathbb R$, then
\begin{gather*}
F=k\left( \frac 1{(x-y)^2}+\frac1{(x-z)^2}+\frac1{(y-z)^2}\right)
\end{gather*}
coincides with the celebrated Calogero potential, which is a well known superintegrable system (see for example~\cite{CDR1} and references therein).
If, with the same choice for the $a_i$,
\begin{gather*}
F=k\left( (X_1+X_2)^{-2}+(2X_1-X_2)^{-2}+(2X_2-X_1)^{-2}\right),
\end{gather*}
then $F$ coincides with the Wolfes potential, a three-body superintegrable interaction whose dynamic equivalence with the Calogero potential is discussed in~\cite{CDR1}. If~$L_0 \neq 0$, then the f\/irst integrals of the extended Hamiltonian for $m=2,3$ and $F$ in the Calogero form  are respectively
\begin{gather*}
U^2G=Gp_u^2-2Aup_uP-4A^2L_0Gu^2,
\\
U^3G=Gp_u^3-3Aup_u^2P-18A^2L_0Gu^2p_u+6A^3L_0u^3P,
\end{gather*}
where $G=(x_1^0+x_2^0+x_3^0+x_1+x_2+x_3)$ and $P=X_L(G)=p_1+p_2+p_3$ is the conserved linear momentum.

\section[Superintegrable extensions of $\mathbb S^2$]{Superintegrable extensions of $\boldsymbol{\mathbb S^2}$}\label{section5}

In \cite{sigma11} the complete solution $G$ of (\ref{HessTeo}) for~$mc=1$  (the constant curvature of a~sphere of radius~1)  is computed in standard spherical coordinates as
\begin{gather}\label{ge}
G= a_0\cos\theta+(a_1\sin\phi+a_2 \cos\phi)\sin\theta.
\end{gather}
Equation (\ref{VTeo}) becomes
\begin{gather*}
\partial_\theta V \partial_\theta G+\frac 1{\sin ^2\theta}\partial_\phi V \partial_\phi G=2VG.
\end{gather*}
Therefore,  $c=\frac 1m\neq 0$ and, by Theorem~\ref{theorem1},   the extension of $L$ is
\begin{gather*}
\frac 12p_u^2+\frac 1{S^2_\kappa(\frac 1mu)}\left(\frac 12 \left(p_\theta^2+\frac 1{\sin^2\theta}p_\phi^2\right)+V\right),
\end{gather*}
where we assume without restrictions the constants $u_0$, $L_0$ and $W_0$ all zero, and where  $\kappa \in \mathbb R$. Since $|\kappa|$ can be multiplied by a positive constant simply by rescaling $u$, we can assume, if $\kappa \neq 0$, $|\kappa|=m^2$ so that the extensions become
\begin{alignat*}{3}
 & H_m^+=\frac 12p_u^2+\frac {m^2}{\sin ^2u}\left(\frac 12 \left(p_\theta^2+\frac 1{\sin^2\theta}p_\phi^2\right)+V\right), \qquad &&\kappa>0, & \\
 & H_m^0=\frac 12p_u^2+\frac {m^2}{u^2}\left(\frac 12 \left(p_\theta^2+\frac 1{\sin^2\theta}p_\phi^2\right)+V\right), \qquad && \kappa=0, & \\
 & H_m^-=\frac 12p_u^2+\frac {m^2}{\sinh ^2u}\left(\frac 12 \left(p_\theta^2+\frac 1{\sin^2\theta}p_\phi^2\right)+V\right), \qquad && \kappa<0. &
\end{alignat*}
For $\kappa \in \mathbb C$, the explicit form of the extension follows from Remark~\ref{remark2}.

\looseness=-1
We remark that  the numerical factor $m^2$ can be absorbed into $L$ by a rescaling of the co\-or\-di\-na\-tes $(\theta,\phi)$. Whenever~$V$ depends on $(\theta,\phi)$ only through trigonometric functions, the rescaling enlights the existence of discrete (polyhedral) symmetries of~$L$ on $\mathbb S^2$ of order depending on~$m$.

The superintegrable potentials on $\mathbb S^2$ with f\/irst integrals all quadratic in the momenta are de\-ter\-mi\-ned in \cite{KM?}, where the sphere is intended, as~$\mathbb E^2$ previously, as a complex manifold. The nine dif\/ferent superintegrable potentials, up to symmetries in $O(3,\mathbb C)$ including ref\/lections,  are, in Cartesian three-dimensional coordinates $(x,y,z)$ with $x=\sin \theta \cos \phi$, $y=\sin \theta \sin \phi$, $z=\cos \theta$~\cite{KM?},
\begin{alignat*}{3}
&S1 \ \ \ && V=\frac {\alpha_1}{\bar w^2}+\frac{\alpha_2 z}{\bar w^3}+\frac {\alpha_3\big(1-4z^2\big)}{\bar w^4},&\\
&S2 \ \ \ && V=\frac {\alpha_1}{z^2}+\frac {\alpha_2}{\bar w^2}+\frac{\alpha_3 w}{\bar w^3},&\\
&S3 \ \ \ && V=\frac {\alpha_1}{z^2},&\\
&S4 \ \ \ && V=\frac {\alpha_1}{\bar w^2}+\frac{\alpha_2 z}{\sqrt{x^2+y^2}}+\frac {\alpha_3}{\bar w\sqrt{x^2+y^2}},&\\
&S5 \ \ \ && V=\frac {\alpha_1}{\bar w^2},&\\
&S6 \ \ \ && V=\frac {\alpha_1 z}{\sqrt{x^2+y^2}},&\\
&S7 \ \ \ && V=\frac {\alpha_1 x}{\sqrt{y^2+z^2}}+\frac {\alpha_2 y}{z^2\sqrt{y^2+z^2}}+\frac {\alpha_3}{z^2},&\\
&S8 \ \ \ && V=\frac {\alpha_1x}{\sqrt{y^2+z^2}}+\frac{\alpha_2(w-z)}{\sqrt{w(z-iy)}}+\frac{\alpha_3(w+z)}{\sqrt{w(z+iy)}},&\\
&S9 \ \ \ && V=\frac {\alpha_1}{x^2}+\frac {\alpha_2}{y^2}+ \frac {\alpha_3}{z^2},&
\end{alignat*}
where $w=x+iy$, $\bar w=x-iy$, $\alpha_1, \alpha_2, \alpha_3 \in \mathbb C$. We remark that in $(x,y,z)$ we have
\begin{gather*}
G=a_2x+a_1y+a_0z,
\end{gather*}
with $a_0,a_1,a_2\in \mathbb C$.

By putting the previous expressions for $V$ and $G$ given by (\ref{ge}) into (\ref{VTeo}), we get a linear homogeneous function in $(a_i)$ whose components are linear homogeneous in $(\alpha_1,\alpha_2,\alpha_3)$.  The non-trivial solutions (with $V$ and $G$  both not constant) of the equation are

\begin{center}
\begin{tabular}{|c|c|c|c|}
\hline
&$V$ & $G$ & particular cases of \\
\hline
$i$ &$\dfrac{\alpha_2}{\bar w^2}+\dfrac{\alpha_3 w}{\bar w^3}$\tsep{5pt}\bsep{9pt} & $a_0z$ & $S2$ \\
$ii$ &$\dfrac{\alpha_1}{\bar w^2}+\dfrac{\alpha_2}{\bar w\sqrt{x^2+y^2}}$\bsep{12pt} & $a_0z$ & $S4$  \\
$iii$ &$\dfrac{\alpha_1}{\bar w^2}$\bsep{8pt} & $a_0z$ & $S1$, $S2$, $S4$, $S5$\\
$iv$ &$\dfrac{\alpha_1}{z^2}+\dfrac{\alpha_2 y}{z^2\sqrt{y^2+z^2}}$\bsep{12pt} & $a_2x$ & $S7$ \\
$v$ &$\dfrac{\alpha_1}{z^2}$ & $a_2x+a_1y$\bsep{8pt} & $S2$, $S3$, $S7$, $S9$ \\
$vi$&$\dfrac{\alpha_1}{x^2}+\dfrac{\alpha_2}{y^2}$\bsep{8pt} & $a_0z$ & $S9$ \\
\hline
\end{tabular}
\end{center}

The only cases without superintegrable extensions for any combination of parameters are~$S6$ and~$S8$, that are strictly related. The extensible cases of $S9$ are all equivalent to~$(v)$ or~$(vi)$ up to permutation of the coordinates. Case~$(v)$ can be considered a subcase of~$(iv)$, but it is listed apart because the corresponding expression of $G$ is dif\/ferent.

\subsection*{Extensions of $\boldsymbol{S9}$}
As an example of the extension procedure, we develop the computations of extensions and f\/irst integrals for the case~$S9$. For the subcase~$(v)$ we have
$V=\frac {\alpha_3}{\cos^2\theta}$,
with
$
G=(a_1\sin\phi+a_2\cos\phi)\sin \theta.
$
For $m=3$  we have for the Hamiltonian $H_3^+$ the f\/irst integral
\begin{gather*}
U^3G = B\sin \theta p_u^3-27B{\rm ctan} ^3u \cos \theta p_\theta ^3-27 C\frac {{\rm ctan}^3 u}{\sin ^3 \theta}p_\phi^3+ B\cos\theta {\rm ctan}\, u \, p_u^2p_\theta\\
\hphantom{U^3G =}{}
+9C\frac{{\rm ctan} u}{\sin \theta}p_u^2p_\phi-27B{\rm ctan} ^2 u\sin \theta p_up_\theta^2
 - 27B\frac {{\rm ctan} ^2  u}{\sin \theta}p_up_\phi^2
-27C\frac{{\rm ctan} ^3 u}{\sin \theta}p_\theta^2 p_\phi
\\
\hphantom{U^3G =}{}
-27B\frac{{\rm ctan} ^3 u \cos \theta}{\sin ^2 \theta} p_\phi^2p_\theta 
 - 54\alpha_3\!\left(\! B\frac {{\rm ctan} ^2 u \sin\theta}{\cos \theta}p_u+B \frac{{\rm ctan}^3  u}{\cos \theta }p_\theta+C\frac{{\rm ctan}^3 u}{\sin\theta \cos^2 \theta}p_\phi\!\right)\!,
\end{gather*}
where $B=a_1\sin \phi+a_2 \cos\phi$, $C=a_1\cos \phi-a_2\sin \phi=\frac {dB}{d\phi}$.

The second of the subcases  of $S9$ admitting a superintegrable extension is $(vi)$
\begin{gather*}
V=\frac 1{\sin^2\theta}\left(\frac {\alpha_1}{\sin^2\phi}+\frac {\alpha_2}{\cos^2\phi}\right),
\end{gather*}
with
$G=\cos \theta.$
For $m=2$  we have for the Hamiltonian $H_2^0$ the f\/irst integral
\begin{gather*}
U^2G=\cos \theta p_u^2-4\frac {\sin \theta}up_\theta p_u-4\frac {\cos \theta}{u^2}p_\theta^2-4\frac{\cos \theta}{u^2\sin ^2\theta}-8\frac{\cos \theta}{u^2}V.
\end{gather*}

\section{Extensions of TTW-type systems}\label{section6}

Let us consider the Hamiltonian
\begin{gather*}
L=\frac 12 p_1^2+\frac \zeta{S_\chi^2(x_1)}\left(\frac12 p_2^2+F(x_2)\right), \qquad \chi, \zeta \in \mathbb R \ \mbox{or} \  \mathbb C.
\end{gather*}
For $\zeta=1$, $\chi$ real and
\begin{gather*}
F(x_2)=\frac{\alpha_1}{\cos^2\lambda x_2}+\frac{\alpha_2}{\sin^2\lambda x_2},
\end{gather*}
$L$ is a generalization to constant curvature manifolds of the Tremblay--Turbiner--Winternitz system
(see \cite{KKM, MPY}). We consider the possible extensions of $L$ in dimension three given by Theorem~\ref{theorem1}. Since the sectional curvature of the metric of~$L$ is $\chi$, for $mc=\chi$ the general complete solution~$G$ of~(\ref{HessTeo}) is
\begin{gather*}
G=a_0C_\chi(x_1)+\left(  a_1S_\zeta(x_2)+a_2C_\zeta(x_2)\right)S_\chi(x_1),
\end{gather*}
and the extended Hamiltonian has the form
\begin{gather*}
H=\frac 12p_u^2+\frac \chi{S_\kappa^2\big(\frac \chi m u\big)}L,
\end{gather*}
where we assume for simplicity $L_0=u_0=W_0=0$.
Equation (\ref{VTeo}) becomes then
\begin{gather}\label{EqF}
 F' \left(a_1C_\zeta(x_2)-a_2\zeta S_\zeta(x_2)\right)=2\zeta F\left(a_1S_\zeta(x_2)+a_2C_\zeta(x_2)\right).
\end{gather}
If $a_1=a_2=0$, then the equation~(\ref{EqF}) is satisf\/ied for all~$F$, including the TTW potential with any value of~$\lambda$. In particular, if $\lambda$ is rational then~$L$ is superintegrable together with its extensions.

Otherwise, if $a_1$ or $a_2$ are dif\/ferent from zero then the solutions $F$ of the equation~(\ref{EqF}) can be obtained after observing that
\begin{gather*}
\zeta\left(a_1S_\zeta(x_2)+a_2C_\zeta(x_2)\right)=-\frac{d}{dx_2}\left(a_1C_\zeta(x_2)-a_2\zeta S_\zeta(x_2)\right).
\end{gather*}
Hence,
\begin{gather*}
F=\frac{1}{\left(a_1C_\zeta(x_2)-a_2\zeta S_\zeta(x_2)\right)^2}.
\end{gather*}
By dif\/ferentiating the relation $a_1S_\zeta(x_2)+a_2C_\zeta(x_2)=\mathcal{A}S_\zeta(x_2+\xi)$, valid for suitable constants~$\mathcal{A}$ and~$\xi$, we have
\begin{gather*}
F=\frac{1}{\mathcal{A}^2C^2_\zeta(x_2+\xi)},
\end{gather*}
a result analogous to the one obtained in \cite{sigma11} for the extension of one-dimensional systems. Indeed,  when $\zeta\in \mathbb N\setminus \{0\}$ then $L$ is  in the form of an extension of a one-dimensional system. This is another example of iterative extension.

\section{Conclusions and future directions}\label{section7}

We have shown how the procedure of extension proposed in \cite{sigma11} can be used to produce new superintegrable systems starting from the already known ones, together with their f\/irst integrals. This procedure allows to extend a number of remarkable systems, including TTW and three-particle Calogero systems. Moreover, in some cases the procedure can be performed iteratively, thus constructing a family of superintegrable systems in higher dimensions.

\looseness=-1
Unfortunately not all the superintegrable systems can be extended through our method, but this drawback is balanced by the simplicity and compactness of the algorithm that produces the constants of motion. Further studies are in progress to f\/ind a more general form of extension  compatible with a larger number of potentials and to analyze iterative extension in other cases. New results about the application to nonconstant curvature manifolds of Theorem~\ref{theorem1} have been obtained. The problem of applying the extension procedure to quantum systems is not yet solved: the f\/irst integrals described by Theorem~\ref{theorem1} cannot be straightforwardly associated with  symmetry operators. However, for the quadratic f\/irst integrals of type $U^2(G)$ a f\/irst quantization procedure has been considered, quite unsuccessfully, in \cite{CDRG} and then a second one has been studied, leading in suitable cases to symmetry operators. All these progresses will be presented in future publications.

\pdfbookmark[1]{References}{ref}
\LastPageEnding


\begin{thebibliography}{99}
\footnotesize\itemsep=0pt

\bibitem{sigma11}
Chanu C., Degiovanni L., Rastelli G., First integrals of extended
  {H}amiltonians in {$n+1$} dimensions generated by powers of an operator,
  \href{http://dx.doi.org/10.3842/SIGMA.2011.038}{\textit{SIGMA}} \textbf{7} (2011), 038, 12~pages, \href{http://arxiv.org/abs/1101.5975}{arXiv:1101.5975}.

\bibitem{CDRG}
Chanu C., Degiovanni L., Rastelli G., Generalizations of a method for
  constructing f\/irst integrals of a class of natural Hamiltonians and some
  remarks about quantization, \href{http://dx.doi.org/10.1088/1742-6596/343/1/012101}{\textit{J.~Phys. Conf. Ser.}} \textbf{343} (2012),
  012101, 15~pages, \href{http://arxiv.org/abs/1111.0030}{arXiv:1111.0030}.

  \bibitem{CDR1}
Chanu C., Degiovanni L., Rastelli G., Superintegrable three-body systems on the
  line, \href{http://dx.doi.org/10.1063/1.3009575}{\textit{J.~Math. Phys.}} \textbf{49} (2008), 112901, 10~pages,
  \href{http://arxiv.org/abs/0802.1353}{arXiv:0802.1353}.


\bibitem{JH}
Jauch J.M., Hill E.L., On the problem of degeneracy in quantum mechanics,
  \href{http://dx.doi.org/10.1103/PhysRev.57.641}{\textit{Phys. Rev.}} \textbf{57} (1940), 641--645.


\bibitem{KM?}
Kalnins E.G., Kress J.M., Pogosyan G.S., Miller Jr. W., Completeness of
  superintegrability in two-dimensional constant-curvature spaces,
  \href{http://dx.doi.org/10.1088/0305-4470/34/22/311}{\textit{J.~Phys.~A: Math. Gen.}} \textbf{34} (2001), 4705--4720,
  \href{http://arxiv.org/abs/math-ph/0102006}{math-ph/0102006}.

\bibitem{KKMCu}
Kalnins E.G., Kress J.M., Miller Jr. W., Talk given by J.~Kress during the
  conference ``Superintegrability, Exact Solvability, and Special Functions''
  (Cuernavaca, February 20--24, 2012).

\bibitem{KKM}
Kalnins E.G., Kress J.M., Miller Jr. W., Tools for verifying classical and
  quantum superintegrability, \href{http://dx.doi.org/10.3842/SIGMA.2010.066}{\textit{SIGMA}} \textbf{6} (2010), 066, 23~pages,
  \href{http://arxiv.org/abs/1006.0864}{arXiv:1006.0864}.


\bibitem{MPY}
Maciejewski A.J., Przybylska M., Yoshida H., Necessary conditions for classical
  super-integrability of a~certain family of potentials in constant curvature
  spaces, \href{http://dx.doi.org/10.1088/1751-8113/43/38/382001}{\textit{J.~Phys.~A: Math. Theor.}} \textbf{43} (2010), 382001,
  15~pages, \href{http://arxiv.org/abs/1004.3854}{arXiv:1004.3854}.

\bibitem{EW}
Rodr{\'{\i}}guez M.A., Tempesta P., Winternitz P., Reduction of superintegrable
  systems: the anisotropic harmonic oscillator, \href{http://dx.doi.org/10.1103/PhysRevE.78.046608}{\textit{Phys. Rev.~E}}
  \textbf{78} (2008), 046608, 6~pages, \href{http://arxiv.org/abs/0807.1047}{arXiv:0807.1047}.

\bibitem{TTW}
Tremblay F., Turbiner A.V., Winternitz P., An inf\/inite family of solvable and
  integrable quantum systems on a plane, \href{http://dx.doi.org/10.1088/1751-8113/42/24/242001}{\textit{J.~Phys.~A: Math. Theor.}}
  \textbf{42} (2009), 242001, 10~pages, \href{http://arxiv.org/abs/0904.0738}{arXiv:0904.0738}.

\end{thebibliography}
\end{document}